\documentclass[12pt]{article}
\usepackage{graphicx}
\begin{document}
\begin{center}
{\LARGE \bf Is there any room for new physics in the muon g-2 problem?}
\footnote{Contribution on International Conference Hadron Structure
'02, September 23.-27., 2002, Herlany, Slovakia}\\
\end{center}

\vspace{10mm}

\begin{center}
{\it E. Barto\v s\footnote{Lab. of Theor. Physics,
JINR, Dubna}}\\

\underline{\it S. Dubni\v cka}\footnote{Inst. of Physics, Slovak
Academy of Sciences, Bratislava, SR},

 {\it A. Z. Dubni\v ckov\'a\footnote{Dept. of Theor. Physics, Comenius University,
Bratislava, SR}},

{\it E.A.Kuraev$^2$ and E.Zemlyanaya$^2$}

\end{center}

\vspace{1cm}

The muon is described by the Dirac equation and its magnetic
moment is related to the spin by means of the expression
\begin{equation}
  \vec{\mu}=g\left({e\over 2m_\mu}\right)\vec{s} \label{z1}
\end{equation}
where the value of gyromagnetic ratio $g$ is predicted (in the
absence of the Pauli term) to be exactly 2.

   However, interactions existing in nature modify $g$
to be exceeding the \mbox{value 2} because of the emission and
absorption of:

\begin{itemize}
\item
 virtual photons (electromagnetic effects),
\item
 intermediate vector and Higgs bosons (weak interaction effects)
\item
 vacuum polarization into virtual hadronic states (strong interaction
effects).
\end{itemize}

In order to describe this modification of $g$ theoretically, the
magnetic anomaly was introduced by the relation

\begin{eqnarray}
  a_\mu\equiv \frac{g-2}{2}&=&a_\mu^{(1)}\left({\alpha\over\pi}
  \right) + \left(a_\mu^{(2)QED}+a_\mu^{(2)had}\right)
  \left({\alpha\over\pi}\right)^2 +  \nonumber \\
  &+& a_\mu^{(2)weak} + O\left({\alpha\over\pi}\right)^3 \label{z2}
\end{eqnarray}
where to every order Feynman diagrams (see Figs. 1-3) correspond
and $\alpha=1/137.03599976(50)$  is the fine structure constant.

\begin{figure}[ht]
\begin{center}
\includegraphics[scale=0.6]{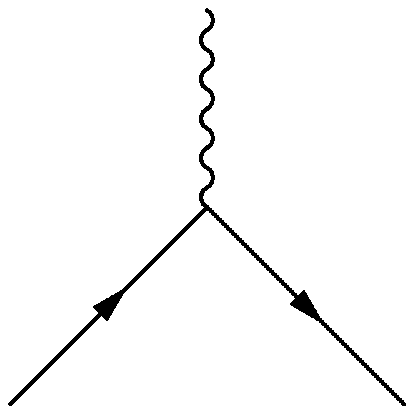}
\caption{The simplest Feynman diagram of an interaction of the
muon with an external magnetic field.} \label{fig:1}
\end{center}
\end{figure}

\begin{figure}[th]
\begin{center}
\includegraphics[scale=0.6]{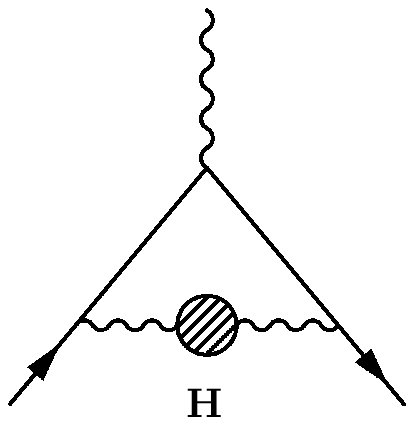}
\caption{The lowest-order hadronic vacuum-polarization
contribution to the anomalous magnetic moment of the muon.}
\label{fig:2}
\end{center}
\end{figure}

\begin{figure}[th]
\begin{center}
\includegraphics[scale=0.6]{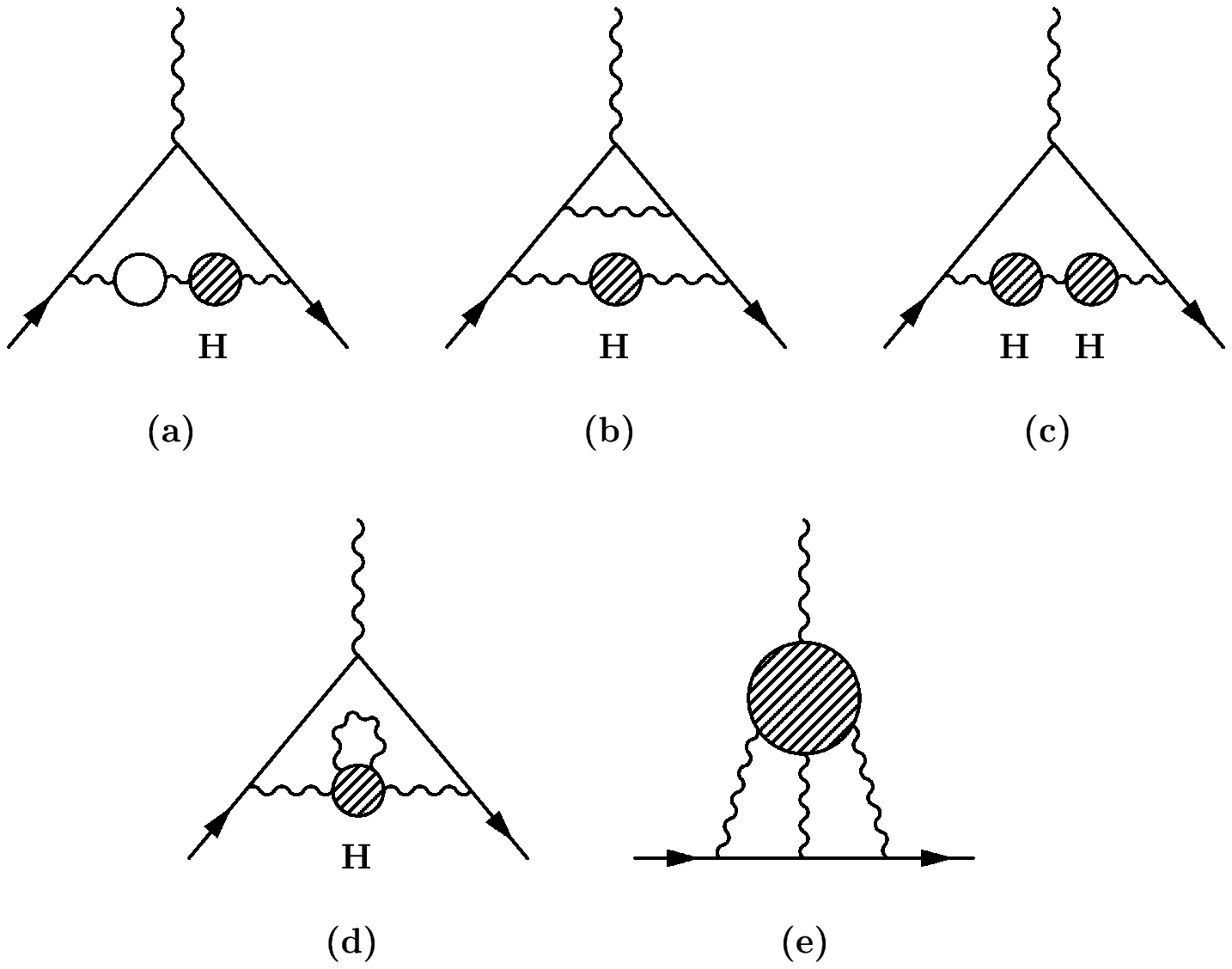}
\caption{The third-order hadronic vacuum-polarization
contributions to the anomalous magnetic moment of the muon.}
\label{fig:3}
\end{center}
\end{figure}

   The muon anomalous magnetic moment $a_{\mu}$ is
very interesting object for theoretical investigations due to
the following reasons:

\begin{enumerate}
  \item[$\left.i\right)$] it is the best measured
   quantity (BNL E--821 experiment) in physics
\begin{equation}
  a_\mu^{exp}=(116 592 040 \pm 86)\times 10^{-11}\cite{[1]} \label{z3}
\end{equation}

  \item[$\left.ii\right)$] its accurate theoretical evaluation
  provides an extremely clean test of "Electroweak theory" and may
  give hints on possible deviations from Standard Model (SM)

  \item[$\left.iii\right)$] moreover, in near future the
  measurement in BNL is expected to be performed yet with an improved
  accuracy
\begin{equation}
  \Delta a_\mu^{exp}=\pm 40\times 10^{-11}\label{z4}
\end{equation}

i.e. it is aimed at obtaining a factor 2 in a precision above
that of the last E--821 measurements.
\end{enumerate}

At the aimed level of the precision (\ref{z4}) a sensibility will
already exist to contributions
\begin{equation}
  a_\mu^{(2,3)weak}=(152 \pm 4)\times 10^{-11},\label{z5}
\end{equation}
arising from single-- and two--loop weak interaction diagrams.
   And so, if we compare theoretical evaluations of:

   QED contributions up to 8th order

\begin{center}
   $a_{\mu}^{QED}=(116 584 705.7 \pm 2.9)\times10^{-11}$\cite{[2]}
\end{center}

   the single- and two-loop weak contributions

\begin{center}
  $a_\mu^{(2,3)weak}=(151 \pm 4)\times 10^{-11}$\cite{[3]}\\

  $a_\mu^{(2,3)weak}=(153 \pm 3)\times 10^{-11}$\cite{[4]}\\

  $a_\mu^{(2,3)weak}=(152 \pm 1)\times 10^{-11}$\cite{[5]}\\
\end{center}

   strong int. contributions

\begin{center}
  $a_\mu^{had}=(7068 \pm 172)\times 10^{-11}$\cite{[6]}\\

  $a_\mu^{had}=(7100 \pm 116)\times 10^{-11}$\cite{[7]}\\

  $a_\mu^{had}=(7052 \pm 76)\times 10^{-11}$\cite{[8]}\\

  $a_\mu^{had}=(7024 \pm 152)\times 10^{-11}$\cite{[9]}\\

  $a_\mu^{had}=(7021 \pm 76)\times 10^{-11}$\cite{[10]}\\
\end{center}
it is straightforward to see that the largest uncertainty is in
$a_{\mu}^{had}$.

   Error is comparable, or in the best case 2x
smaller than the weak interaction contributions.

   So, in order to test the SM predictions for
$a_{\mu}$ and to look for new physics in comparison with BNL
E--821 experiment, one has still to improve an evaluation of
$a_{\mu}^{had}$.

   The most critical from all hadronic contributions are the
light--by--light (LBL) meson pole terms (see Fig.4) and we have
recalculated them in the paper\cite{[11]}.

\begin{figure}[th]
\begin{center}
\includegraphics[scale=0.6]{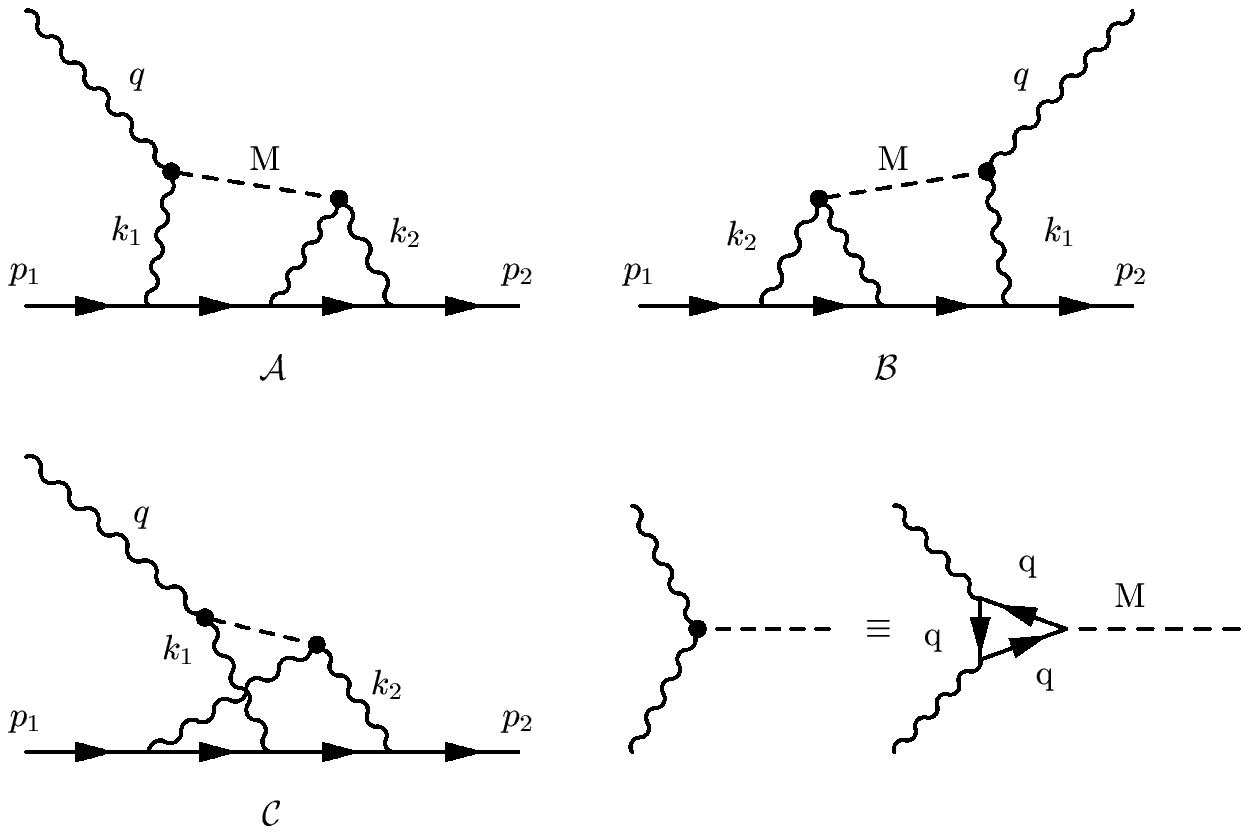}
\caption{Meson (M) pole diagrams in the third order hadronic
light--by--light scattering contributions to $a_\mu^{had}$.}
\label{fig:4}
\end{center}
\end{figure}

More concretely we have evaluated contributions of the scalar
$\sigma$, $a_0$ and pseudoscalar $\pi^0$, $\eta$, $\eta'$ mesons
$(M)$ in the framework of the linearized extended Nambu--Jona--Lasinio
model

\begin{center}
$\mathcal{L}$$_{q\bar qM}= g_M \bar q(x)\left[\sigma(x)+i \pi(x)
\gamma_5\right]q(x).$
\end{center}

   The reason for the latter are predictions of series of recent
papers

\begin{center}
   $a_\mu^{LBL}=(+52 \pm 18) \times 10^{-11}$\cite{[12]}\\

   $a_\mu^{LBL}=(+92 \pm 32) \times 10^{-11}$\cite{[13]}\\

   $a_\mu^{LBL}=(+79.2 \pm 15.4) \times 10^{-11}$\cite{[14]}\\

   $a_\mu^{LBL}=(+83 \pm 12) \times 10^{-11}$\cite{[15]}\\

   $a_\mu^{LBL}(\pi_0)=(+58 \pm 10)\times 10^{-11}$\cite{[16]}\\
\end{center}

which differ in the magnitude.

   Moreover, in these papers only the pseudoscalar pole contributions
were considered.

   We include the scalar meson $(\sigma, a_0)$ pole contributions as
well.

   Current methods in a description of the  $\gamma^*\rightarrow M\gamma^*$
transition form factors are ChPT and the vector--meson--dominance
(VMD) model.

   Here the corresponding transition form
factors by the constituent quark triangle loops with
colourless and flavourless quarks with charge equal to the
electron one are represented.

   An application of a similar modified constituent quark triangle
loop model for a prediction of the pion electromagnetic form
factor behaviour was carried out in \cite{[17]} where also a
comparison with the naive VMD model prediction was demonstrated.

   The mass of the quark in the triangle loop is taken to be:

\begin{center}
$m_u =m_d=m_q= (280 \pm 20)$ MeV\\
\end{center}
determined \cite{[18]} in the framework of the chiral quark model
of the Nambu--Jona--Lasinio type by exploiting the experimental
values of the pion decay constant, the $\rho$-meson decay into
two-pions constant, the masses of pion and kaon and the mass
difference of $\eta$ and $\eta '$ mesons.

   The unknown strong coupling constants of $\pi^0,\eta,\eta '$ and
$a_0$ mesons with quarks are evaluated in a comparison of the
corresponding theoretical two-photon widths with experimental
ones.

   The $\sigma$-meson coupling constant is taken to be equal to
$\pi^0$-meson coupling constant as it follows from the
corresponding Lagrangian.

   The $\sigma$-meson mass is taken to be $m_\sigma$=$(496 \pm 47)$
MeV as an average of the values recently obtained experimentally
from the decay $D^+ \rightarrow \pi^-\pi^+\pi^+$ \cite{[19]} and
excited $\Upsilon$ decay \cite{[20]} processes.

   As a result we present explicit formulas for $a_\mu^{LBL}(M)$
$(M=\pi^0, \eta, \eta ', \sigma, a_0)$ in terms of Feynman
parametric integrals of 10-dimensional order, which subsequently
are calculated by MIKOR method.

As a result one finds
\begin{eqnarray}
     a_\mu^{LBL}(\pi^0) &=& (81.83 \pm 16.50) \times 10^{-11}\nonumber\\
     a_\mu^{LBL}(\eta) &=& (5.62 \pm 1.25) \times 10^{-11}\nonumber\\
     a_\mu^{LBL}(\eta ') &=& (8.00 \pm 1.74) \times 10^{-11} \\
     a_\mu^{LBL}(\sigma) &=& (11.67 \pm 2.38) \times 10^{-11}\nonumber\\
     a_\mu^{LBL}(a_0) &=& (0.62 \pm 0.24) \times 10^{-11}.\nonumber
\end{eqnarray}

   So, the total contribution of meson poles in LBL is

\begin{equation}
             a_\mu^{LBL}(M) = (107.74 \pm 16.81) \times 10^{-11},
\end{equation}
where the resultant error is the addition in quadrature of all
partial errors of (6).

   Together with the contributions of the pseudoscalar meson $(\pi^\pm,
K^\pm)$ square loops and constituent quark square loops (Fig.5)
taken from Hayakawa and Bijnens it gives

\begin{figure}[ht]
\begin{center}
\includegraphics[scale=0.6]{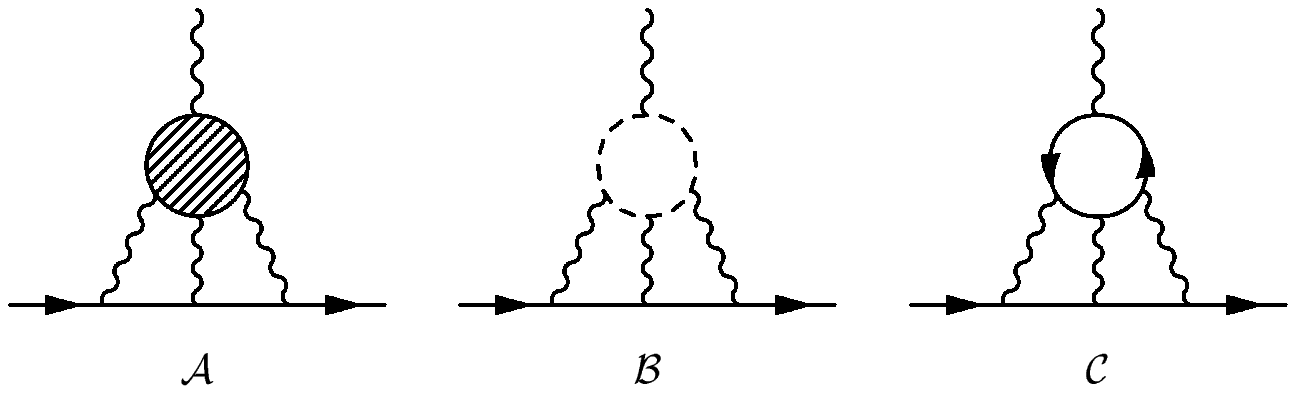}
\caption{Third order hadronic light--by--light scattering
contribution to $a_\mu^{had}$ ($\mathcal{A}$) and class of
pseudoscalar meson square loop diagrams ($\mathcal{B}$) and quark
square loop diagrams ($\mathcal{C}$) contributing to
($\mathcal{A}$).} \label{fig:5}
\end{center}
\end{figure}

\begin{equation}
            a_\mu^{LBL}(total) = (111.20 \pm 16.81) \times 10^{-11}.
\end{equation}

   The others 3-loop hadronic contributions derived from the hadronic
vacuum polarizations $(VP)$ were most recently evaluated by
Krause\cite{[21]}

\begin{equation}
              a_\mu^{(3)VP} = (-101 \pm 6) \times 10^{-11}.
\end{equation}

   Then the total 3-loop hadronic correction is

\begin{equation}
       a_\mu^{(3)had} = a_\mu^{LBL}(total) + a_\mu^{(3)VP} =
                        (10.20 \pm 17.28) \times 10^{-11}
\end{equation}
where the errors have been again added in quadratures.

   If we take into account the most recent evaluation \cite{[22]}
of the lowest--order hadronic vacuum--polarization contribution to the
anomalous magnetic moment of the muon

\begin{equation}
              a_\mu^{(2)had} = (7021 \pm 76) \times 10^{-11}
\end{equation}
the pure QED contribution up to 8th order

\begin{equation}
              a_\mu^{QED} = (116 584 705.7 \pm 2.9) \times 10^{-11}
\end{equation}
and the single-- and two--loop weak interaction contribution,
finally one gets the SM theoretical prediction of the
muon anomalous magnetic moment value to be

\begin{equation}
              a_\mu^{th} = (116 591 888.9 \pm 78.1) \times10^{-11}.
\end{equation}

   Comparing this theoretical result with experimental one finds

\begin{equation}
              a_\mu^{exp} - a_\mu^{th} = (151 \pm 116) \times 10^{-11}
\end{equation}
which implies a reasonable consistency of the SM prediction for
the anomalous magnetic moment of the muon with experiment.

   However, one expects in near future a 2x lowering of the error in BNL E--821
experiment and then there can appear still a room for a new
physics beyond the SM.

   On the other hand, we see some possible improvements for theoretical
value, which as a result could diminish the difference between theoreticaly
estimated value and the measured value in E--827 experiment.

   The first improvements we see still in the lowest-order hadronic
vacuum-polarization diagram contributions (Fig.2), which can be expressed
by the integral

\begin{equation}
a_\mu^{(2) had}=\frac{1}{4\pi^3}\int_{4m_\pi^2}^{\infty}
\sigma^{tot}(s) K_\mu(s) ds;
\end{equation}
where  $\sigma^h(s)$ stands for the total cross section $\sigma(e^+e^-\to had)$
and
\begin{equation}
K_\mu(s)=\int_0^1\frac{x^2(1-x)}{x^2+(1-x)s/m_\mu^2} dx.
\end{equation}

   We have indications, that the $e^+e^-\rightarrow K^+K^-$ and
$e^+e^-\rightarrow K^0\bar K^0$ data at the $\phi$-resonance
region, measured at the Novosibirsk, are inconsistent with
analyticity.

   They have to be systematically shifted to the larger values and so
they will give larger positive contributions approaching the
theoretically estimated value to the experimental one.

   The same can be said about the contribution of the processes
$e^+e^-\rightarrow \pi \gamma$, $e^+e^-\rightarrow \eta \gamma$
and $e^+e^-\rightarrow \eta'\gamma$, which were not estimated up
to now as there is unknown a behaviour of corresponding transition
form factors in the time-like region.

   We have elaborated the unitary and analytic model which solves the
latter problem.

   The last contribution which according to our knowledge was not considered
up to now is $K^0$ meson square loop diagram in the third order
hadronic light-by-light scattering contribution to
$a_{\mu}^{had}$.

   We hope to realize all these ideas before obtaining the final result
at the BNL E--821 experiment with the precision

  $\Delta a_\mu^{exp}=\pm 40\times 10^{-11}$.

\end{document}